\documentclass[a4paper,10pt]{article}

\usepackage{graphicx,setspace}

\usepackage{amsmath,textcomp}

\begin{document}

\title{The dynamic nature of conflict in Wikipedia} 

\author{Y. Gandica, F. Sampaio dos Aidos \and J. Carvalho \\
\emph{Centre for Computational Physics, Department of Physics, }\\
\emph{University of Coimbra, 3004-516 Coimbra, Portugal} }


\date{\today}

\maketitle
\singlespacing
\begin{abstract}
The voluntary process of Wikipedia edition provides an environment where the outcome is clearly a 
collective product of interactions involving a large number of people. 
We propose a simple agent-based model, developed from real data, to reproduce the collaborative process of Wikipedia edition. 
With a small number of simple ingredients, our model mimics several interesting features of real human behaviour,
namely in the context of edit wars. 
We show that the level of conflict is determined by a tolerance 
parameter, which measures the editors' capability to accept different opinions and to change their own opinion. 
We propose to measure conflict with a parameter based on mutual reverts, which increases only in contentious situations. 
Using this parameter, we find a distribution for the inter-peace periods that is heavy-tailed. 
The effects of wiki-robots in the conflict levels and in the edition patterns are also studied.
Our findings are compared with previous parameters used to measure conflicts in edit wars.
\end{abstract}

\footnotetext{$^{1}$Correspondence author. E-mail: ygandica@gmail.com}

\newpage \baselineskip1.0cm
\singlespacing
\section{Introduction}

The study of interacting particle systems has, for a long time, been an important 
subject of Physics. The use of statistical methods has allowed for major advances in this area, by 
providing a bridge between the microscopic interactions and the large collective behaviour of the system~\cite{huang,kardar}. 
This success has motivated researchers to try a statistical approach to other subjects outside Physics \cite{rmp_haken,book_barrat}. 
The application of the methods of Statistical Physics to social phenomena, where the interacting 
particles are now interacting human beings, has proved to be very fruitful in allowing for the understanding 
of many features of human behaviour \cite{rmp_barabasi,rmp_dorogovtsev,rmp_castellano}. 
Some of these properties are common to very different phenomena in nature. Scaling, for example, is 
generally observed in a great variety of human networks. Universality, which states that the emergent phenomena displayed by the collective 
behaviour of interacting particles depends on symmetries, dimensionality and conservation laws and not on
the microscopic details of the intrinsic dynamics mechanism \cite{kardar,goldenfeld}, seems to be present 
in many social situations \cite{rmp_barabasi,rmp_dorogovtsev,rmp_castellano}. 
In this sense, the cornerstone to the successful modelling of social systems 
depends mainly on two major strategies: on one hand an appropriate selection of the relevant variables and,
on the other, a good visualisation/representation of the displayed phenomena, both related to the specific system being studied.

For this endeavour, Internet data has played an important role as a source of data that allows 
for the test of models of universal social patterns as a collective effect of interaction among single individuals \cite{book_newman}. 
The development of an article for Wikipedia (WP) is a process in which anybody may edit and change its
content. 
Whatever the reasons behind someone's decision to edit an article in WP, and whatever his
background and previous knowledge on the article's subject, it has been recognised that the 
reliability thus obtained is comparable to that of other high quality professional encyclopedias, 
such as the Encyclopedia Britannica \cite{nature_quality_wikipedia}. 
This voluntary process clearly creates an environment where the outcome is a collective product of interactions among a large number of
people \cite{wikipedia_as_networks,visual_analisis,assesing_the_value}. 
The online availability of the historical record of all editions has promoted an intense research activity 
\cite{review1,review2,razones_para_controversia}, trying to grasp the intrinsic behaviour that 
characterises several features of WP editing.

The emergence and development of conflicts in WP is one of the features that has recently received 
attention from the academic world. 
The interplay between strong convictions and tolerance leads to 
conflicts among the editors (the so-called edit wars) and the WP article may converge (or not) 
into a consensus edition \cite{indentiying_controversy}. 
Several approaches to measure the level of conflict in an article as it develops have been tried. 
Some take into account the resulting topology, for example
on semantic flow \cite{semantic_wikipedia} or on talk pages \cite{12030652v1}, missing the underlying dynamical process. 
Other approaches determine conflict levels by focusing on WP editors' 
dynamical behaviour, by measuring, for example, the talk page length \cite{030407,talk_pages,talk_pages2}.
The drawback of these approaches, besides the time consuming effort needed to apply the methodologies, is
that the use of this editing channel, that was created to discuss changes and controversies, depends on cultural traits. 
In some cultures, the talk pages are extensively used to discuss the 
differences of opinion, while in others they are almost not used at all for this purpose \cite{anterior1}. 
The search for culture independent 
indicators led to the study of disputes between pairs of editors by measuring, for example, the number of 
words that they have deleted from each other in a sequence of editions \cite{indentiying_controversy,ranking_controversy}. 
These and other appropriate parameters, that show some degree of mutual correlation, have been shown to 
work well as measures of conflict \cite{anterior1}. 

Recently, it was realised that, in the context of edit wars, reversion is quite common and becomes a 
typical mechanism used by editors to disagree with others in a controversial mode \cite{revert_network}.
Reversion consists of completely recovering a certain previous edition of the article, totally 
disregarding the changes made afterwards. 
However, revert maps cannot fully differentiate between 
conflictive and non-conflictive articles \cite{anterior1}, as reversion cannot discriminate between dispute 
and the response to simple acts of vandalism (such as restoring a page that has been fully deleted). 
Furthermore, acts of vandalism are not such a rare event; they are actually responsible for 
about 24\% of all the reverts \cite{revert_network}. 
Nonetheless, this problem can be avoided if, instead of simple reverts, mutual reverts (when two editors revert each other's 
editions) are used to define conflictual behaviours~\cite{primer_mutual,anterior5_value_production,anterior1}. Yasseri {\it et al.} proposed a parameter which is a function of mutual reverts and may
distinguish conflicts from mere vandalism (see \cite{anterior5_value_production} and references 
therein). They also proposed an agent-based model in order to reproduce the main features depicted by 
their controversy parameter \cite{anterior2}.

In this letter, we address the behaviour underlying the collective dynamics that emerges from 
WP editors' interactions. We show some plots of WP data different from those previously presented in the 
literature. The analysis of these plots led us to propose a new agent-based model to simulate the edition 
of a WP page. Inspired by the parameter M introduced by Yasseri et al. \cite{anterior1} to measure 
conflict, we define another parameter C, also based on mutual reverts, which is similar to M but has 
the advantage of more accurately detecting the end of conflicts. We use the same conflict parameter to 
compare our model to real data.
We find a scaling behaviour when measuring the inter-peace periods with the new conflict parameter. 
Finally, we explore how the results of WP data analysis change due to the presence of edit robots. 

\section{ Real data.}  The analysis presented in this letter is based on the January 2010 dump of the English 
WP, containing 4.64 million pages, available at the WikiWarMonitor webpage (http://wwm.phy.bme.hu/). 
The data sample used, a ``light dump", contains a reduced information list of all the pages edits (the edit timestamp, 
a reversal flag, the edition number and the editor identification). 
Only pages with more than $1800$ editions and a lifetime over $6$ years were further analysed.

\section{ Editor's activity.} 
\begin{figure}[h]
\scalebox{0.4}{\includegraphics{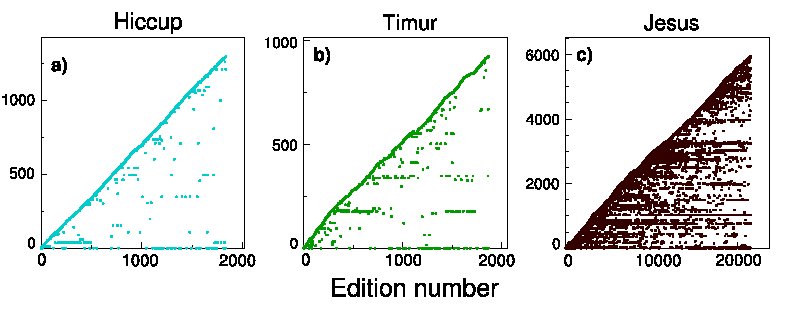}} \caption{Examples of editor's activity as a function of edition number, a) in blue for a 
low, b) in green for a medium and c) in brown for a high controversy page. 
On the vertical axis each editor is numbered according to the chronological order of his debut in that article.}
\label{fig1}
\end{figure}

In order to study the editors' activity, the editions of the article are 
numbered in chronological order. The edition number is denoted by $e$, so that $e=1$ is the first edition, when the article is created, and 
editions $e=2,3,4,...$ are the subsequent updates. 
In fig. \ref{fig1}, we present examples of editors' 
activity as a function of edition number (each editor is numbered according to the chronological order of
his debut in that particular article), for a low, a medium and a high controversy page, as defined later.
A symbol is plotted in each graph for every edition of every editor. 
In the three cases shown in this 
figure, there is a ratio $R$ between the number of editors and the number of editions that is approximately constant 
over time. We found a similar behaviour in all the articles that we checked, and this seems to be a reasonable assumption. 

\begin{figure}[h]
\scalebox{0.4}{\includegraphics{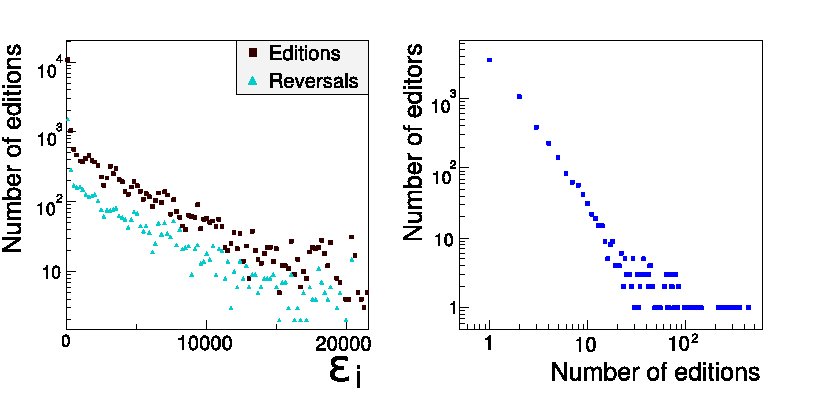}} \caption{For the wikipage ``Jesus". 
Left panel: In brown, the number of editions in each bin, sized $\sim 200$ editions, as a function of $\varepsilon_i$, summed over all editors. 
In blue, the same for reversal editions. 
Right panel: distribution of the number of editions per editor.}
\label{fig2}
\end{figure}

It is also clear from fig. \ref{fig1} that the activity of most editors decays over time. 
We highlight this effect by summing all editions made by all editors after they start to edit. Let $e_i$ be the 
edition number at which editor $E_i$ edits the article for the first time. Then, $\varepsilon_i=e-e_i$ is the 
edition number after $E_i$ starts to edit. 
We choose the article named ``Jesus" (with a total of 21,768 
editions) and plot in fig. \ref{fig2} the editing activity as a function of $\varepsilon_i$. 
For that purpose, we divide the total number of editions in 100 bins, each of size $\Delta \varepsilon$ equal to just
over 200 editions, and plot the sum of the number of editions in each bin for all editors on the left
panel in brown (for example, the first dot in this figure is the total number of times that all editors
have edited the article during the first $\Delta\varepsilon$ editions since they started to edit). 
The brown dots follow approximately a straight line in the semi-log graph, which suggests an exponential decay in editing
activity for the average editor. 
The number of reversal editions is shown in blue in the same graph, following a similar pattern. 
On the right panel of fig. \ref{fig2} we plot the distribution of the number of editors as a function of 
the total number of editions that each one has made, for the same article. 
The distribution seems to
follow a straight line again, but now in a log-log graph, suggesting a power law distribution.

\section{ Agent-based model.}  We propose an agent-based model that tries to grasp the features observed in real WP page edition. Like Yasseri {\it et al.} \cite {anterior5_value_production}, we use the Bounded Confidence model proposed by Deffuant {\it et al.} \cite{deffuant} as a suitable candidate to address the dynamics of collaboration and conflicts in WP edition. We denote by  $x_i$ the value of the opinion of editor $E_i$ and by $A$ the state of the article, i.e., the opinion reflected by the article, where an opinion about the subject under discussion is quantified by a continuous value between 0 and 1. Several sets of initial opinion distributions were tried, such as a uniform random distribution between 0 and 1 or a combination of Gaussian distributions with different parameters. We found the final results to be qualitatively unaffected by this choice and we have used for the initial opinion of the editors a Gaussian distribution with mean $m=0.5$ and standard deviation $\sigma=0.1$, with a cutoff 
for values less than zero or greater than one. The computer simulation starts with just one editor and, at each dynamical step $e$, a new editor comes in and edits the wikipage for the first time with probability $R$. This means that, at each time step, an old editor $E_i$ will interact with the article with probability $1-R$, in which case the probability for choosing $E_i$ among all the available editors will be:
\begin{equation}
P(E_i) = \frac{\eta_i N_i}{\sum_j \eta_j N_j}
\label{eq1}\end{equation}
where $N_i$ is the number of previous editions by editor $E_i$ and $\eta_i$ is a random number between 0 and 1. 
The proportionality to $N_i$ is similar to the preferential attachment effect 
\cite{barabasi_albert}, ``edits beget edits", in Wilkinson and Huberman's words \cite{assesing_the_value}.
Bryant {\it et al.} showed that the involvement of the editors with the quality of a WP article increases
with the number of times they have edited it\cite{wikipedian}. 
We assume this increased involvement may be described by the proportionality to $N_i$ in eq.(\ref{eq1}) \cite{anterior5_value_production,visual_analisis}. 
The parameter $\eta_i$ is intended to measure the editor's propensity to edit this particular article, either due to the extent 
of his knowledge on the subject or merely to some emotional connection to it (the fitness as
defined by Bianconi and Barab\'asi in \cite{bianconi_barabasi}). Once $\eta_i$ is defined for editor $E_i$, it will maintain its value throughout the whole edition procedure.
The sum is performed over all active editors.

After choosing which editor is going to interact with the article in a specific dynamical step, the 
edition process is decided as follows: if the difference between the editor's opinion, $x_i$, and the
article's current state, $A$, is inside a tolerance threshold, $\epsilon$, ($|x_i-A|<\epsilon$),
the editor will change his mind and approach the article's point of view, which remains unchanged, 
$ \Delta x_i = -\mu (x_i-A)$ where $\mu (=0.2)$ is a convergence parameter. 
Otherwise, the editor will maintain his opinion and change the article. 
In the latter case, if a previous edition is found 
with an opinion value difference from the current editor's opinion smaller than $\epsilon$, then with 
probability $P_{rev}(=0.5)$, the editor will revert the article to a previous (the nearest) such edition. 
In case no edition reversal occurs, the article state is changed to approach the editor's opinion, 
$ \varDelta A = \mu (x_i-A)$. 
After each edition, each active editor will become inactive with probability $P_{inac} (=0.0005)$, which reflects 
the editor's loss of interest in the article (an ageing effect). 
This parameter tries to mimic the observed
progressive loss of interest of most editors in the article, which can be perceived in fig. \ref{fig1}.
We have chosen the reversal probability value according to our findings about reversals in data, 
as we show, for example, on the left panel of fig.~\ref{fig2}, for the specific case of the wikipage ``Jesus". 
The value is able to reproduce the reversal behaviour in pages of different controversiality.

\begin{figure}[h]
\scalebox{0.4}{\includegraphics{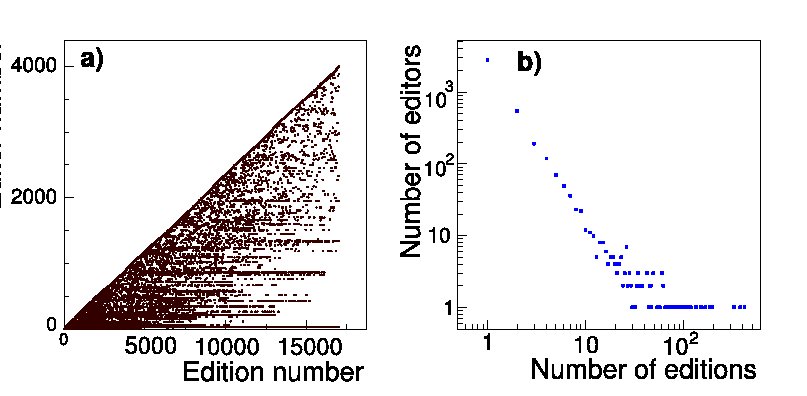}} \caption{ From our simulation model: a) Editors' activity as a function of edition number ($R \approx 0.24$), 
b) Distribution of the number of editions per editor.}
\label{fig3}
\end{figure}

In fig. \ref{fig3} we show the simulation results for 17k editions and 4k editors. 
Plot a) shows the editor's activity (similar to fig. \ref{fig1} c) for real data). 
Plot b) shows the distribution of the number of editions per editor (similar to the right panel of fig. 
\ref{fig2} for real data). In summary, our model has five parameters, three of which have the 
following fixed values for all simulations: $P_{inac} =0.0005$, $\mu=0.2$ and $P_{rev} =0.5$, which were 
obtained so as to reproduce the real data. The parameter $R$ is adjusted to describe the amount of editor
participation required for each simulation and $\epsilon$ is the only parameter that controls the degree 
of controversy.

\section{ Controversy parameters.}  Several algorithms have been proposed to rank controversial articles in WP; most of them concentrate 
in detecting edit wars and high controversiality  \cite{anterior1}. 
Yasseri {\it et al.} measured controversy by means of a sum over the minimum number of all the editions by each pair of editors, with 
at least one mutual reversal between them
\begin{equation}
M=E\times\sum_{(N_i^d,N_j^r)<\max}\min(N_i^d,N_j^r)
\label{eq2}\end{equation}
where $\min(N_i^d,N_j^r)$ is the minimum of the 2 values $N_i^d$ and $N_j^r$, which are the number of editions of editors $E_i$ and $E_j$ 
who have been involved in at least a mutual reversal with each other and $E$ is the number of editors who, at some point, have performed a mutual reversal 
with any other editor. 
The sum excludes the term with the maximum value, in order to remove a possible personal conflict.

Yasseri {\it et al.} \cite{anterior1} showed that this parameter can effectively select the high controversy WP articles. 
However, there is a problem with this definition. 
Assuming there is a collaborative period after a conflict, this parameter keeps growing, as long as the editors involved in the conflict 
keep editing, thus failing to recognise the end of the conflict.

In order to avert this problem, we reworked $M$ and define our conflict parameter, $C$, as the sum of all the reversals between all pairs 
of editors with at least a mutual reversal between them $\langle i,j \rangle$, multiplied by $E$ (keeping the general recognition that 
``the larger the armies, the larger the war"),
\begin{equation}
C = E \times \sum_{\langle i,j \rangle} N_{i,j}^R
\label{eq3}\end{equation}
where $N_{i,j}^R$ is the number of reversals between editors $E_i$ and $E_j$ (both the reversals of $E_i$ over an edition by $E_j$ and vice versa).
We chose not to exclude the maximum in this sum, as there is no way to identify a personal conflict (except by actually looking 
at the several editions of the WP page in detail). 

\begin{figure}[h]
\scalebox{0.4}{\includegraphics{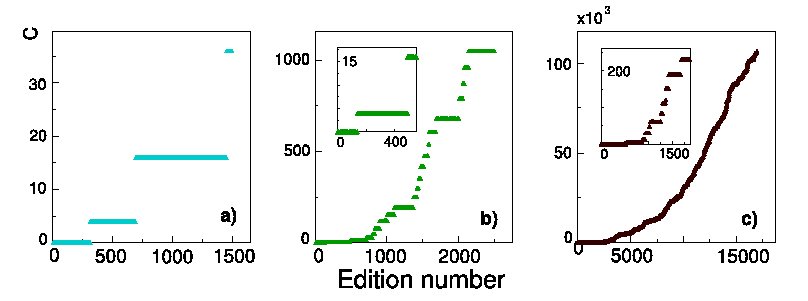}} \caption{ From our simulation model: controversy parameter ($C$) as a function of edition number for a) $ \epsilon=0.18$ (low controversy, blue), b) $\epsilon=0.10$ (medium controversy, green) and c) $\epsilon=0.05$ (high controversy, brown). 
Insets of the medium and high controversy plots show zooms with behaviours similar to low and medium controversy, respectively (a scale effect).}
\label{fig3a}
\end{figure}

\begin{figure}[h]
\scalebox{0.4}{\includegraphics{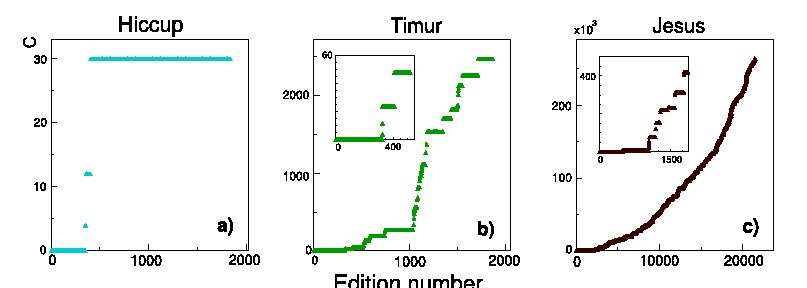}} \caption{Controversy parameter $C$ as a function of edition number for WP pages of a) low (``Hiccup"), b)
medium (``Timur") and c) high controversy (``Jesus"). 
Again, insets of the medium and high controversy plots show zooms with behaviours similar to low and medium controversy, respectively.}
\label{fig4}
\end{figure}

Fig. \ref{fig3a} displays the conflict parameter $C$ as a function of the edition number for the same simulation and for different values of the tolerance parameter and number of editions. In plot a) we show the evolution of $C$ for $\epsilon=0.18$, in plot b) for $\epsilon=0.10$ and in plot c) for $\epsilon=0.05$, which correspond, respectively, 
to a low, a medium and a high level of conflict. 
Comparison with plots obtained with real data (shown in fig.~\ref{fig4}), suggest that the tolerance parameter provides an 
appropriate description of the level of conflict in the edition process.

\begin{figure}[h]
\scalebox{0.4}{\includegraphics{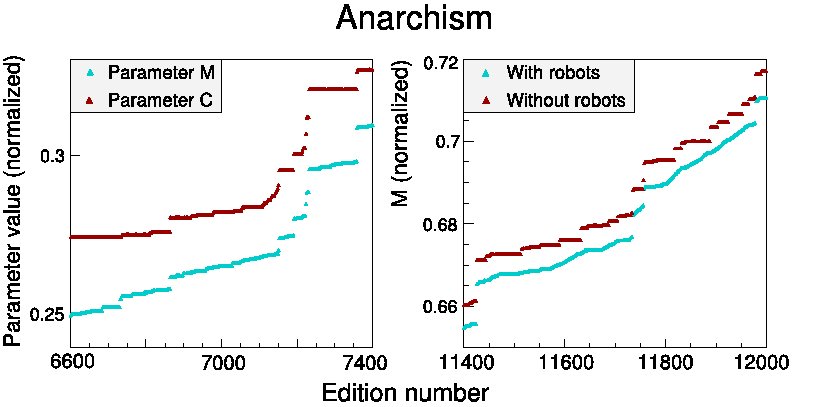}} \caption{Left panel: An example of normalised values of $C$ (brown) and $M$ (blue) 
for the same editing interval. Note how M rises when C remains constant in several intervals, like at 
the beginning of the plot.
Right panel: An example of $M$ evolution with (blue) and without (brown) robots \cite{bots}, for the same 
period. Both plots were obtained with the wikipage "Anarchism".}
\label{fig5}
\end{figure}

In identifying the highly controversial pages, parameters $C$ and $M$ are quite similar. 
Out of the top 100 most controversial pages according to each of those two parameters, 
$ 80 \%$ are common to both selections, and the measured correlation between the two parameters is $r=0.97$~\cite{correlation}. 
Out of the 4.64 million pages of the English WP, only about 216k have $C>0$ and, out of these about 58\% (just over 124k) have $C\leq 10$. 
For medium and high controversy pages, $\sim 5.8$k (2.7\%) have $C>1$k and $\sim 650$ (0.3\%) have $C>10$k.
Despite the above mentioned similarities,
the left panel of fig. \ref{fig5}, showing normalised values of $C$ and $M$ for the same edition period, illustrates that the two parameters 
follow different evolutions. 
There are some peaceful periods (where $C$ is constant) that are not recognised as such by $M$ (which increases most of the time).
$M$ is suitable to detect the most controversial articles, but it fails in capturing some collaboration patterns. 

It is important to recognize that the edition of a WP page is shared between humans and robots, i.e., programs conceived to perform 
various routine tasks such as spell correction and vandalism detection. 
In order to measure conflict activity among humans and uncover the intrinsic dynamics of controversy, we must remove the 
effect of the robots in the controversy parameters \cite{bots}. 
On the right panel of fig. \ref{fig5} we compare the evolution of parameter $M$ with and without the robots. 
It can be seen that the robots do make a difference in conflict detection, as 
their non elimination artificially increases the conflict parameter $M$. A similar effect is not observed in parameter $C$, where the difference between the two graphs (with and without robots) is not perceivable with the naked eye (and for this reason we do not show it here).

There is a significant correlation between high levels of conflict and the number of editions \cite{anterior5_value_production}. 
Therefore, when comparing conflictuality periods, it may be necessary to normalise the measuring parameter with respect to the number of editions.

\section{ Inter-peace periods.} 
\begin{figure}[h]
\scalebox{0.4}{\includegraphics{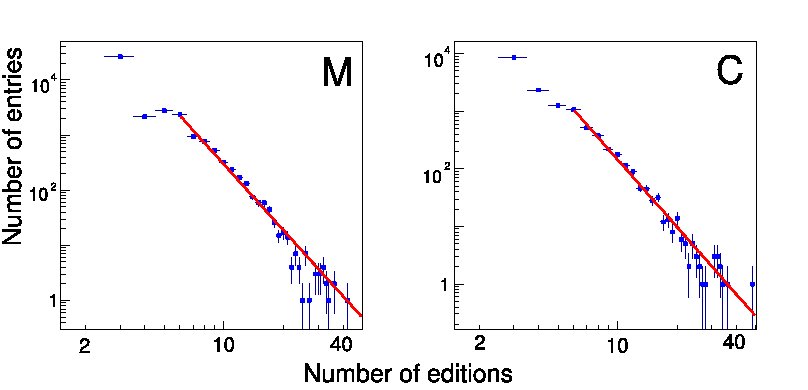}} \caption{Distribution of the number of editions between peaceful intervals for $M$ and $C$ on the left and 
on the right panel, respectively. The red lines are power law fits to the data points.}
\label{fig6}
\end{figure}

A lot of effort has been devoted to describe and understand temporal patterns of human 
activities \cite{burst_nature_barabasi,how_to_burst}. 
Several heavy-tailed distributions of time intervals between events on different communication media have been found. 
WP has shown specific time attributes depending on geographical and cultural constraints 
\cite{anterior5_value_production, temporal_geografical_patterns,detecting_vandalism_west}. 
Yasseri {\it et al.} reported the exponent of the inter-edit time distribution \cite{temporal_geografical_patterns}. 
In order to analyse edition patterns of edit wars, we show in fig.~\ref{fig6} the fat-tailed distribution of inter-peace 
periods, where we define peace as a period, lasting no less than $n$ editions, during which the controversy parameter remains constant 
\cite{data}. In our calculations, we have used $n=3$ and we have gathered all the pages with more than 1500
editions and a lifetime longer than 8 years. 
We believe that this makes more sense than to study the length of the peace periods as these will, in many cases, 
depend on exogenous factors either cyclic (like the anniversary of some event related to the article) or non cyclic 
(special events) \cite{fluctuations_wikipedia,time_evolution_wikipedia}. 

The red line in fig \ref{fig6} represents our power law fit to the data. With parameter $C$, we
get for the exponent the value $\alpha=3.90\pm 0.06$, with a fit $p$-value of 0.012, while for $M$ we get
an almost negligibly higher $\alpha$ value ($3.98\pm 0.04$), because it does not take into account some of
the peaceful times between conflictual editors. The $p$-value for this fit is practically zero 
($\sim 2\times 10^{-35}$). The details of these calculations are explained in refs 
\cite{especificaciones_fit,xmin}. The fit $p$-value was obtained from the $\chi^2$ fit probability.

The robots were excluded from all these calculations. We checked that their effect on the values of $\alpha$ is very small.

\section{ Conclusions.} The edit wars on WP are conflicts with the unusual circumstance of being a symmetric scenery, in which
the antagonist entities are individuals with no connection to one another. 
In this letter, we aimed at capturing the basic ingredients that lead the WP editing process to a collaborative/consensual 
edition or to remain conflictive. 
With a simple agent-based model that relies on few parameters, we reproduced several interesting real behaviours. 
We proposed a conflict measurement parameter based on mutual reverts that is culture independent and has the advantage, 
over previous parameters, of being able to differentiate between collaboration and conflict among editors. 
We show that the level of conflict in a WP page can be related to the tolerance in the system, which may be associated to issues that people are not 
prepared to negotiate because of the strong convictions involved. 
We found a long-tailed power law distribution on inter-peace periods, measured with our conflict measuring parameter, $C$. 
We also showed the differences in the power law exponent with a previous conflict parameter, $M$. 
We mention the effect of the robots that edit the WP in the controversy measurement parameter and edit patterns. 

This study has not included the real edition time. 
The parameter $e$ was defined as the edition number and all variables 
were defined in terms of $e$ and not of the real time. 
A natural extension of this work will be to study the dynamical features in terms of real time.

JC thanks Filipe Veloso and Cl\'audio Silva for help on technical questions. 
Y.G. Thanks Ernesto Medina and Silvia Chiacchiera for useful discussions .
YG acknowledges financial support from Fundos FEDER through Programa Operacional Factores de 
Competitividade-COMPETE and by Funda\c{c}\~{a}o para a Ci\^{e}ncia e a Tecnologia, through Project FCOMP-01-0124-FEDER-015708.


\begin{thebibliography}{10}

\bibitem{huang} 
Huang K. Statistical Mechanics. Massachussetes Institute of Technology. 1963.

\bibitem{kardar}
Kardar M. Statistical Physics of Particles. Cambridge University. Press. 2007.

\bibitem{rmp_haken}
Haken H. Rev. Mod. Phys. 47 (1975) 67.

\bibitem{book_barrat}
Barrat A., Barth\'{e}lemy M. and Vespignani. A. Dynamical Processes on Complex Networks.
(Cambridge) 2008.

\bibitem{rmp_barabasi}
Albert R. and Barab\'{a}si A.~L. Rev. Mod. Phys. 74 (2002) 47.

\bibitem{rmp_dorogovtsev}
Dorogovtsev S., Goltsev A. and Mendes J. Rev. Mod. Phys. 80 (2008) 1275.

\bibitem{rmp_castellano}
Castellano C., Fortunato S. and Loreto V. Rev. Mod. Phys. 81 (2009) 591.

\bibitem{goldenfeld}
Goldenfeld N. Lectures on Phase Transitions and the Renormalization Group.
Perseus Books Publishing, L.L.C. 1992.

\bibitem{book_newman}
Newman M., Barabasi A.-L. and Watts. D. The Structure and Dynamics of NETWORKS. 
(Princeton University Press) 2006.

\bibitem{nature_quality_wikipedia}
Giles J. Nature. 438 (2005) 900.

\bibitem{wikipedia_as_networks}
V~Zlati\'{c}, M~Bo\v{z}i\v{c}evi\'{c} H.~v. and Domazet M. Phys. Rev. E. 74 (2006) 016115.

\bibitem{visual_analisis}
Brandes U. and Lerner J. Visual analysis of controversy in user-generated encyclopedias.
http://dx.doi.org/10.1145/1391107.1391111 (2007).

\bibitem{assesing_the_value}
Wilkinson D.~M. and Huberman. B.~A. Information Society Watch, First Monday. 12 (2007) 4.

\bibitem{review1}
Okoli C., Mehdi M., Mesgari M., Nielsen F.~A. and Lanam\"{a}ki A. 
The people’s encyclopedia under the gaze of the sages: A systematic
review of scholarly research on wikipedia. http://dx.doi.org/10.2139/ssrn.2021326 (2012).

\bibitem{review2}
Nicolas J. What we know about wikipedia. A review of the literature analyzing the project(s).
http://dx.doi.org/10.2139/ssrn.2053597 (2012).

\bibitem{razones_para_controversia}
Li C., Datta A. and Sun A. Social Netw. Analys. Mining. 2(3) (2012) 265.

\bibitem{indentiying_controversy}
Rad H.~S. and Barbosa D. Identifying controversial articles in wikipedia: A comparative study.
http://dx.doi.org/10.1145/2462932.2462942 (2012).

\bibitem{semantic_wikipedia}
Masucci A.~P., Kalampokis A., Eguiluz V.~M. and Hern\'{a}ndez-Garc\'{i}a. E.
PLoS ONE 6, Iss. 2 (2011) e17333.

\bibitem{12030652v1}
Vicent~G\'{o}mez, Hilbert J.~Kappen N.~L. and Kaltenbrunner A. World Wide Web. 16 (2013) 645–675.

\bibitem{030407}
Vi\'{e}gas F.~B., Wattenberg M., Kriss J. and van Ham F. Talk before you type: Coordination in
wikipedia. Http://dx.doi.org/10.1109/HICSS.2007.511 (2007).

\bibitem{talk_pages}
Schneider J., Passant A. and Breslin J.~G. A content analysis: How wikipedia talk pages are 
used. Http://hdl.handle.net/10379/2501 (2012).

\bibitem{talk_pages2}
Laniado D., Tasso R., Volkovich Y. and Kaltenbrunner A. When the wikipedians talk: Network and 
tree structure of wikipedia discussion pages. 5th International AAAI Conference on Weblogs and Social
Media, ICWSM 2011, pp. 177-184 (2011).

\bibitem{anterior1}
Yasseri T., Sumi R., Rung A., Kornai A. and Kert\'{e}sz J. PLos One. 7, Iss. 6 (2012) e38869.

\bibitem{ranking_controversy}
Vuong B.-Q., Lim E.-P., Sun A., Le M.-T., Lauw H.~W. and Chang K. On ranking controversies in 
wikipedia: Models and evaluation. ISBN: 978-1-59593-927-2 (2008).

\bibitem{revert_network}
Suh B., Chi E.~H., Pendleton B.~A. and Kittur A. Us vs. them: Understanding social 
dynamics in wikipedia with revert graph visualizations. http://dx.doi.org/10.1109/VAST.2007.4389010 (2007).

\bibitem{primer_mutual}
Sumi R., Yasseri T., Rung A., Kornai A. and Kert\'{e}sz J. Characterization and prediction of wikipedia 
edit wars. J. Proceedings of the ACM WebSci11, Koblenz, Germany, pp. 1-3 (2011).

\bibitem{anterior5_value_production}
Yasseri T. and Kert\'{e}sz J. J Stat Phys. 151 (2013) 414.

\bibitem{anterior2}
T\"or\"ok J., Iniguez G., Yasseri T., San-Miguel M., Kaski K. and Kert\'{e}sz J.
PRL 110 (2013) 088701.

\bibitem{deffuant}
Deffuant G., Neau D., Amblard F. and Weisbuch G. Adv. Compl. Syst 03 (2000) 87.

\bibitem{barabasi_albert}
Barab\'{a}si A.-L. and Albert R. SCIENCE 286 (1999) 509.

\bibitem{wikipedian}
Bryant S.~L., Forte A. and Bruckman A. Becoming wikipedian: Transformation of 
participation in a collaborative online encyclopedia. http://dx.doi.org/10.1145/1099203.1099205 (2005).

\bibitem{bianconi_barabasi}
Bianconi G. and Barab\'{a}si A.-L. Europhys. Lett 54 (4) (2001) 436.

\bibitem{correlation}
The correlation value $r$ was obtained using the Pearson's product-moment
  coefficient, as defined at athworld.wolfram.com/CorrelationCoefficient.html.
  A similar result is obtained with the Spearman rank correlation coefficient.

\bibitem{bots}
List of active bots in \\ en.wikipedia.org/wiki/Category:Active\_Wikipedia\_bots.

\bibitem{burst_nature_barabasi}
Barab\'{a}si A.-L. Nature 435 (2005) 207.

\bibitem{how_to_burst}
Goh K. and Barab\'{a}si A.-L. Europhys. Lett. 81 (2008) 48002.

\bibitem{temporal_geografical_patterns}
Yasseri T., Sumi R. and Kert\'{e}sz J. PLOS ONE 7 (2012) e30091.

\bibitem{detecting_vandalism_west}
West A.~G., Kannan S. and Lee. I. Detecting wikipedia vandalism via spatio-temporal analysis of 
revision metadata. http://dx.doi.org/10.1145/1752046.1752050 (2010).

\bibitem{fluctuations_wikipedia}
K\"{a}mpf M., Tismer S., Kantelhardt J.~W. and Muchnik. L. Physica A 391 (2012) 6101.

\bibitem{time_evolution_wikipedia}
Kaltenbrunner A. and Laniado D. There is no deadline - time evolution of wikipedia discussions.
http://dx.doi.org/10.1145/2462932.2462941 (2012).

\bibitem{especificaciones_fit}
For the calculation of the exponent in the fat-tailed distributions we
  compared the results according to the procedure in {\sc Clauset A., Rohilla
  Shalizi C.} and {\sc Newman M. E. J.}. SIAM REVIEW. Vol. 51, No. 4, pp.
  661703 (2009) and the ones obtained with the software ROOT ({ \sc Brun R}.
  and {\sc Rademakers F.}, Proceedings AIHENP'96 Workshop, Lausanne, Sep. 1996,
  Nucl. Inst. \& Meth. in Phys. Res. A 389 (1997) 81-86. See also
  http://root.cern.ch/.http://root.cern.ch/drupal).

\bibitem{xmin}
The Kolmogorov-Smirnov test was used for the calculation of $x_{min}$ (lower fit range), 
according with the procedure described in the previous reference.

\end{thebibliography}
\end{document}